\documentclass[conference]{IEEEtran} 
\IEEEoverridecommandlockouts
\usepackage{cite}
\usepackage{amsmath,amssymb,amsfonts}
\usepackage{algorithmic}
\usepackage{graphicx}
\usepackage{textcomp}
\usepackage{xcolor}
\usepackage{verbatim}
\def\BibTeX{{\rm B\kern-.05em{\sc i\kern-.025em b}\kern-.08em
    T\kern-.1667em\lower.7ex\hbox{E}\kern-.125emX}}
\begin{document}

\title{Functional Connectivity of Imagined Speech and Visual Imagery based on Spectral Dynamics\\
{\footnotesize \textsuperscript{ }}
\thanks{20xx IEEE. Personal use of this material is permitted. Permission
from IEEE must be obtained for all other uses, in any current or future media, including reprinting/republishing this material for advertising or promotional purposes, creating new collective works, for resale or redistribution to servers or lists, or reuse of any copyrighted component of this work in other works.

This work was supported by the Institute for Information \& Communications Technology Promotion (IITP) grant funded by the Government of South Korea (No. 2017-0-00451, Development of BCI-based Brain and Cognitive Computing Technology for Recognizing User’s Intentions using Deep Learning; No. 2019-0-00079, Artificial Intelligence Graduate School Program, Korea University).}
}


\author{\IEEEauthorblockN{Seo-Hyun Lee}
\IEEEauthorblockA{\textit{Dept. Brain and Cognitive Engineering} \\
\textit{Korea University}\\
Seoul, Republic of Korea \\
seohyunlee@korea.ac.kr}
\and
\IEEEauthorblockN{Minji Lee}
\IEEEauthorblockA{\textit{Dept. Brain and Cognitive Engineering} \\
\textit{Korea University}\\
Seoul, Republic of Korea \\
minjilee@korea.ac.kr}
\and
\IEEEauthorblockN{Seong-Whan Lee}
\IEEEauthorblockA{\textit{Dept. Artificial Intelligence} \\
\textit{Korea University}\\
Seoul, Republic of Korea \\
sw.lee@korea.ac.kr}
}

\maketitle

\begin{abstract}
Recent advances in brain-computer interface technology have shown the potential of imagined speech and visual imagery as a robust paradigm for intuitive brain-computer interface communication. However, the internal dynamics of the two paradigms along with their intrinsic features haven’t been revealed. In this paper, we investigated the functional connectivity of the two paradigms, considering various frequency ranges. The dataset of sixteen subjects performing thirteen-class imagined speech and visual imagery were used for the analysis. The phase-locking value of imagined speech and visual imagery was analyzed in seven cortical regions with four frequency ranges. We compared the functional connectivity of imagined speech and visual imagery with the resting state to investigate the brain alterations during the imagery. The phase-locking value in the whole brain region exhibited a significant decrease during both imagined speech and visual imagery. Broca and Wernicke’s area along with the auditory cortex mainly exhibited a significant decrease in the imagined speech, and the prefrontal cortex and the auditory cortex have shown a significant decrease in the visual imagery paradigm. Further investigation on the brain connectivity along with the decoding performance of the two paradigms may play a crucial role as a performance predictor.
\end{abstract}

\begin{IEEEkeywords}
\textit{electroencephalography; functional connectivity; imagined speech; intuitive brain-computer interface; visual imagery}
\end{IEEEkeywords}

\section{Introduction}

Recently, imagined speech and visual imagery paradigms are actively investigated in the field of intuitive brain-computer interface (BCI). These endogenous paradigms are reported to be effective in BCI communication since they involve the user intention directly  \cite{b1}. Attempts to robust decoding of the two paradigms with the investigation of their intrinsic features are being actively proposed \cite{b2}. However, the underlying features and cortical networks of imagined speech and visual imagery are still missing. Understanding the intrinsic cortical networks may play a crucial role in improving the decoding performance of BCI paradigms. Yet, the intrinsic features and the cortical networks have not been taken into account with the classification performance. Considering the intrinsic features of the two paradigms may highly improve the decoding performance \cite{n01}.

Previous studies have demonstrated the possibility of decoding imagined speech in terms of phonemes, words, and simple sentences phonemes \cite{b3}, words \cite{b4}, and simple sentences \cite{b5}. For multiclass decoding, studies reported three-class classification performance of 50.1\% using Riemannian features and a relevance vector machine \cite{b6} and 54.1\% using a discrete wavelet transform feature with a multilayer perceptron \cite{b7}. Qureshi \textit{et al}. \cite{b8} reported 32.9\% of the five-class imagined speech classification accuracy using hybrid connectivity features and an extreme learning machine. As for visual imagery, the state-of-the-art performance remains at lower levels—55.9\% for binary decoding \cite{b9} and 25.9\% for six-class classification \cite{b10}. Lee \textit{et al}. \cite{b1} reported 20.4\% and 22.2\% in thirteen-class classification of imagined speech and visual imagery, respectively. These performances may further be improved by considering the intrinsic features of each paradigm.

For the intrinsic features and brain dynamics, the left posterior superior temporal cortex (Wernicke’s area) and the left prefrontal and premotor regions, including Broca area and the supplementary motor area (SMA), are known to be involved in the imagined speech process \cite{b11}. Fronto-temporal coherence is also demonstrated as a measure of connectivity in speech control \cite{b12}. Additionally, it has been reported for decreased brain activity during imagery \cite{b6}, high activation on the temporal area \cite{b8}, and high-frequency activation \cite{b13}. The increased top-down information flow in parieto-occipital cortices is introduced as a distinct feature of visual imagery \cite{b14}. It was reported that visual imagery draws most of the same neural machinery as visual perception \cite{b15}. Visual imagery is known to be associated with the activation of the temporal and occipital regions \cite{b15}.

In the present study, we investigated the brain dynamics of imagined speech and visual imagery paradigm by comparing the brain connectivity of imagery and resting state in the two paradigms. We analyzed the functional connectivity using the phase-locking value (PLV) in specific frequency ranges along with various cortical regions. By focusing on the brain state alterations during imagery of the two paradigms, we aim to find out which features of the brain dynamics highly represent the two intuitive paradigms, imagined speech and visual imagery. Considering the brain dynamics of the two paradigms may contribute to the precise analysis of electroencephalography (EEG) data, thereby enhancing the decoding performance \cite{b16}.

\section{Methods}
\subsection{Data Acquisition}
We used the EEG dataset of sixteen subjects performing the imagery of thirteen-class imagined speech and visual imagery words/phrases. The experimental protocol followed the previous work by Lee \textit{et al}. \cite{b1}. The thirteen-class included twelve words/phrases (ambulance, clock, hello, help me, light, pain, stop, thank you, toilet, TV, water, and yes) and the resting state. The data were recorded via Brain Vision/Recorder (BrainProduct GmbH, Germany) with a 64-channel EEG cap following the international 10-10 system (reference: FCz; ground: FPz). Each subject randomly performed 88 trials of imagined speech and visual imagery per each class (2000 ms was provided per trial). For the imagined speech session, subjects were instructed to silently pronounce the given word without moving the articulators; subjects imagined the given visual scene of each class in the visual imagery session. Subjects were instructed to relax during the rest class. The study was conducted under the Declaration of Helsinki. The experimental protocols were reviewed and approved by the Institutional Review Board at Korea University [KUIRB-2019-0143-01]. All the subjects signed an informed consent form.

\subsection{EEG Data Processing}
The pre-processing of the EEG data was performed using the EEGLAB toolbox \cite{b17}. We down-sampled the data in 256 Hz, and bandpass filtered into 0.5-125 Hz. 60 Hz line noise was removed by applying the notch filter. The common average reference was used for re-referencing the data \cite{b18}. 2000 ms epochs were extracted with the baseline corrected with -500 ms from the onset. The data were then filtered into four frequency ranges (delta: 0.5-4 Hz; theta: 4-8 Hz; alpha: 8-13 Hz; beta: 13-30 Hz). The ocular and muscular artifacts were removed using the automatic algorithms with second-order blind identification and blind source canonical correlation analysis \cite{b19}.

\subsection{Functional Connectivity}
Brain alteration while performing the two paradigms was identified for a precise understanding of the feasibility of brain decoding. Brain connectivity was analyzed using the PLV, which is a measure of the synchronization of electrical brain activities for functional connectivity \cite{b20}. PLV was computed during the imagery and resting state in four frequency ranges within the seven cortical regions (six cortical groups + 64-channel group) as:
\begin{equation}
PLV_{t,i,k}=\displaystyle{\frac{i}{N}|\sum_{n=1}^{N}\exp{(j\theta_{i,k}(t,n))}|}\label{eq}   
\end{equation}
\begin{equation}
\displaystyle{\theta_{i,k}(t,n) = \phi_i(t,n) - \phi_k(t,n);}\label{eq}   
\end{equation}
where $\theta_{i,k}(t,n)$ is the phase difference between channels \textit{i} and \textit{k} in trials \textit{n}; and \textit{N} is the number of trials \cite{b20}. Since PLV is a measure between two channels, we used the grand averaged value of every combination of two channels involved in each cortical region. Inter-cortical connectivity was also evaluated using the averaged PLV of every combination between different cortical groups. Six cortical groups were the Broca and Wernicke’s areas (AF3, F3, F5, FC3, FC5, T7, C5, TP7, CP5, and P5), visual cortex (POz, PO3, PO4, PO7, PO8, PO9, PO10, Oz, O1, and O2), auditory cortex (FT7, FT8, FT9, FT10, T7, T8, TP7, TP8, TP9, and TP10), the motor cortex (Fz, F1, F2, F3, F4, FC1, FC2, FC3, FC4, and Cz), prefrontal cortex (Fp1, Fp2, AF3, AF4, AF7, AF8, F5, F6, F7, and F8), and sensory cortex (Cz, C1, C2, C3, C4, CPz, CP1, CP2, CP3, and CP4) \cite{b21}. The results were verified with statistical analysis. A paired t-test was performed to compare the PLV of imagery versus resting-state EEG.

\begin{table}[b!] 
\caption{Comparison Between PLV of Imagined Speech and Resting State\\(0.5-4 Hz Frequency Range)}
\resizebox{\columnwidth}{!}{
\begin{tabular}{|c|c|c|c|c|}
\hline
\textbf{} & \textbf{\textit{Imagery}} & \textbf{\textit{Rest}} & \textbf{\textit{t-value}} & \textbf{\textit{p-value}}\\
\hline
Whole brain & \textbf{0.32} & \textbf{0.33} & \textbf{-2.892} & \textbf{0.011} \\ \hline
Broca and Wernicke's area & \textbf{0.35} & \textbf{0.37} & \textbf{-2.600} & \textbf{0.020} \\ \hline
Visual cortex & 0.57 & 0.55 & 1.272 & 0.223 \\ \hline
Auditory cortex & 0.33 & 0.34 & -1.500 & 0.154 \\ \hline
Motor cortex & 0.51 & 0.49 & 1.354 & 0.196 \\ \hline
Prefrontal cortex & 0.53 & 0.54 & -1.156 & 0.266 \\ \hline
Sensory cortex & 0.46 & 0.44 & 1.707 & 0.108 \\ \hline
\multicolumn{5}{l}{Significant values are indicated in bold (\textit{p}${<}$0.05)}
\end{tabular}
}
\label{tab1}
\end{table}

\begin{table}[b!] 
\caption{Comparison Between PLV of Imagined Speech and Resting State\\(4-8 Hz Frequency Range)}
\resizebox{\columnwidth}{!}{
\begin{tabular}{|c|c|c|c|c|}
\hline
\textbf{} & \textbf{\textit{Imagery}} & \textbf{\textit{Rest}} & \textbf{\textit{t-value}} & \textbf{\textit{p-value}}\\
\hline
Whole brain & \textbf{0.32} & \textbf{0.34} & \textbf{-4.663} & \textbf{${<}$ 0.001} \\ \hline
Broca and Wernicke's area & \textbf{0.39} & \textbf{0.40} & \textbf{-2.457} & \textbf{0.027} \\ \hline
Visual cortex & 0.55 & 0.55 & -0.604 & 0.555 \\ \hline
Auditory cortex & 0.37 & 0.38 & -2.081 & 0.055 \\ \hline
Motor cortex & 0.58 & 0.57 & 0.488 & 0.633 \\ \hline
Prefrontal cortex & 0.49 & 0.50 & -1.672 & 0.115 \\ \hline
Sensory cortex & 0.46 & 0.46 & ${<}$0.001 & 1.000 \\ \hline
\multicolumn{5}{l}{Significant values are indicated in bold (\textit{p}${<}$0.05)}
\end{tabular}
}
\label{tab1}
\end{table}

\begin{table}[t!] 
\caption{Comparison Between PLV of Imagined Speech and Resting State\\(8-13 Hz Frequency Range)}
\resizebox{\columnwidth}{!}{
\begin{tabular}{|c|c|c|c|c|}
\hline
\textbf{} & \textbf{\textit{Imagery}} & \textbf{\textit{Rest}} & \textbf{\textit{t-value}} & \textbf{\textit{p-value}}\\
\hline
Whole brain & 0.37 & 0.37 & -0.387 & 0.704 \\ \hline
Broca and Wernicke's area & 0.45 & 0.44 & 0.214 & 0.833 \\ \hline
Visual cortex & 0.55 & 0.55 & ${<}$0.001 & 1.000 \\ \hline
Auditory cortex & 0.38 & 0.38 & 0.822 & 0.424 \\ \hline
Motor cortex & 0.64 & 0.64 & 1.202 & 0.248 \\ \hline
Prefrontal cortex & 0.48 & 0.49 & -1.635 & 0.123 \\ \hline
Sensory cortex & \textbf{0.56} & \textbf{0.54} & \textbf{4.208} & \textbf{0.001} \\ \hline
\multicolumn{5}{l}{Significant values are indicated in bold (\textit{p}${<}$0.05)}
\end{tabular}
}
\label{tab1}
\end{table}

\begin{table}[t!] 
\caption{Comparison Between PLV of Imagined Speech and Resting State\\(13-30 Hz Frequency Range)}
\resizebox{\columnwidth}{!}{
\begin{tabular}{|c|c|c|c|c|}
\hline
\textbf{} & \textbf{\textit{Imagery}} & \textbf{\textit{Rest}} & \textbf{\textit{t-value}} & \textbf{\textit{p-value}}\\
\hline
Whole brain & \textbf{0.30} & \textbf{0.32} & \textbf{-4.243} & \textbf{0.001} \\ \hline
Broca and Wernicke's area & 0.39 & 0.39 & -1.499 & 0.155 \\ \hline
Visual cortex & 0.50 & 0.51 & -0.182 & 0.858 \\ \hline
Auditory cortex & \textbf{0.33} & \textbf{0.34} & \textbf{-2.683} & \textbf{0.017} \\ \hline
Motor cortex & 0.54 & 0.53 & 0.802 & 0.435 \\ \hline
Prefrontal cortex & 0.46 & 0.47 & -1.370 & 0.191 \\ \hline
Sensory cortex & 0.44 & 0.44 & ${<}$0.001 & 1.000 \\ \hline
\multicolumn{5}{l}{Significant values are indicated in bold (\textit{p}${<}$0.05)}
\end{tabular}
}
\label{tab1}
\end{table}

\begin{table}[t!] 
\caption{Comparison Between PLV of Visual Imagery and Resting State\\(0.5-4 Hz Frequency Range)}
\resizebox{\columnwidth}{!}{
\begin{tabular}{|c|c|c|c|c|}
\hline
\textbf{} & \textbf{\textit{Imagery}} & \textbf{\textit{Rest}} & \textbf{\textit{t-value}} & \textbf{\textit{p-value}}\\
\hline
Whole brain & \textbf{0.33} & \textbf{0.36} & \textbf{-2.324} & \textbf{0.035} \\ \hline
Broca and Wernicke's area & 0.36 & 0.38 & -1.773 & 0.097 \\ \hline
Visual cortex & 0.60 & 0.61 & -0.653 & 0.524 \\ \hline
Auditory cortex & 0.36 & 0.37 & -0.972 & 0.347 \\ \hline
Motor cortex & 0.53 & 0.55 & -0.874 & 0.396 \\ \hline
Prefrontal cortex & 0.53 & 0.55 & -1.553 & 0.141 \\ \hline
Sensory cortex & 0.46 & 0.48 & -1.217 & 0.242 \\ \hline
\multicolumn{5}{l}{Significant values are indicated in bold (\textit{p}${<}$0.05)}
\end{tabular}
}
\label{tab1}
\end{table}

\section{Results}
\subsection{Brain State Alteration During Imagery}
We compared the PLV of the imagery with that of the resting state in each cortical region for both paradigms (Tables I-VIII). The PLV of imagined speech significantly decreased in the whole brain region in the delta, theta, and beta bands (Table I, II, and IV). Broca and Wernicke’s areas displayed a significant decrease in PLV of imagined speech in the delta and theta frequency (Table I, II); a significant decrease of PLV of imagined speech was found in the auditory cortex in the beta band range (Table IV). In contrast, the PLV of imagined speech significantly increased in the sensory cortex in the alpha band (Table III). 
For visual imagery, the PLV significantly decreased in comparison with the resting state in the whole-brain region in all four frequency regions (Table V-VIII). Theta band displayed significantly decreased PLV of the visual imagery in the auditory cortex and the prefrontal cortex regions. Additionally, the auditory cortex exhibited a significant decrease in the PLV of visual imagery, compared to the resting state.

\subsection{PLV of Different Cortical Regions}
As shown in Tables I–VIII, a significant decrease in the PLV of imagery state in comparison with the resting state was mainly found in the cortical regions with inferior PLV than the other regions. For instance, the PLV of imagined speech in the whole-brain region had shown the most inferior value compared to any other cortical regions in the delta (0.32), theta (0.32), and beta band (0.30). Also, the PLV of visual imagery exhibited the lowest value in the whole brain region in delta (0.33), theta (0.33), alpha (0.37), and beta (0.31) groups. All of the above groups exhibited a significant decrease in PLV of imagery compared to the resting state. On the contrary, the PLV of imagined speech in the sensory cortex with alpha-band exhibited the highest value (0.56) among the PLV of other cortical groups. It exhibited a significant increase in PLV during imagery compared to the resting state.

\begin{table}[t!] 
\caption{Comparison Between PLV of Visual Imagery and Resting State\\(4-8 Hz Frequency Range)}
\resizebox{\columnwidth}{!}{
\begin{tabular}{|c|c|c|c|c|}
\hline
\textbf{} & \textbf{\textit{Imagery}} & \textbf{\textit{Rest}} & \textbf{\textit{t-value}} & \textbf{\textit{p-value}}\\
\hline
Whole brain & \textbf{0.33} & \textbf{0.36} & \textbf{-2.419} & \textbf{0.029} \\ \hline
Broca and Wernicke's area & 0.41 & 0.42 & -1.414 & 0.178 \\ \hline
Visual cortex & 0.59 & 0.59 & 0.110 & 0.914 \\ \hline
Auditory cortex & \textbf{0.39} & \textbf{0.41} & \textbf{-2.284} & \textbf{0.037} \\ \hline
Motor cortex & 0.60 & 0.61 & -1.072 & 0.300 \\ \hline
Prefrontal cortex & \textbf{0.50} & \textbf{0.52} & \textbf{-2.643} & \textbf{0.018} \\ \hline
Sensory cortex & 0.45 & 0.46 & -1.269 & 0.224 \\ \hline
\multicolumn{5}{l}{Significant values are indicated in bold (\textit{p}${<}$0.05)}
\end{tabular}
}
\label{tab1}
\end{table}

\begin{table}[t!] 
\caption{Comparison Between PLV of Visual Imagery and Resting State\\(8-13 Hz Frequency Range)}
\resizebox{\columnwidth}{!}{
\begin{tabular}{|c|c|c|c|c|}
\hline
\textbf{} & \textbf{\textit{Imagery}} & \textbf{\textit{Rest}} & \textbf{\textit{t-value}} & \textbf{\textit{p-value}}\\
\hline
Whole brain & \textbf{0.37} & \textbf{0.40} & \textbf{-2.387} & \textbf{0.031} \\ \hline
Broca and Wernicke's area & 0.46 & 0.47 & -1.817 & 0.089 \\ \hline
Visual cortex & 0.58 & 0.59 & -1.806 & 0.091 \\ \hline
Auditory cortex & 0.40 & 0.41 & -1.059 & 0.306 \\ \hline
Motor cortex & 0.66 & 0.67 & -1.766 & 0.098 \\ \hline
Prefrontal cortex & 0.49 & 0.51 & -2.017 & 0.062 \\ \hline
Sensory cortex & 0.57 & 0.57 & -0.346 & 0.734 \\ \hline
\multicolumn{5}{l}{Significant values are indicated in bold (\textit{p}${<}$0.05)}
\end{tabular}
}
\label{tab1}
\end{table}

\begin{table}[t!] 
\caption{Comparison Between PLV of Visual Imagery and Resting State\\(13-30 Hz Frequency Range)}
\resizebox{\columnwidth}{!}{
\begin{tabular}{|c|c|c|c|c|}
\hline
\textbf{} & \textbf{\textit{Imagery}} & \textbf{\textit{Rest}} & \textbf{\textit{t-value}} & \textbf{\textit{p-value}}\\
\hline
Whole brain & \textbf{0.31} & \textbf{0.34} & \textbf{-2.470} & \textbf{0.026} \\ \hline
Broca and Wernicke's area & 0.39 & 0.41 & -1.936 & 0.072 \\ \hline
Visual cortex & 0.52 & 0.53 & -0.629 & 0.539 \\ \hline
Auditory cortex & \textbf{0.34} & \textbf{0.36} & \textbf{-2.469} & \textbf{0.026} \\ \hline
Motor cortex & 0.56 & 0.57 & -1.742 & 0.102 \\ \hline
Prefrontal cortex & \textbf{0.46} & \textbf{0.48} & \textbf{-2.337} & \textbf{0.034} \\ \hline
Sensory cortex & 0.44 & 0.45 & -1.115 & 0.282 \\ \hline
\multicolumn{5}{l}{Significant values are indicated in bold (\textit{p}${<}$0.05)}
\end{tabular}
}
\label{tab1}
\end{table}

\begin{figure*}[t!]
\includegraphics[width = \textwidth, height = 9cm]{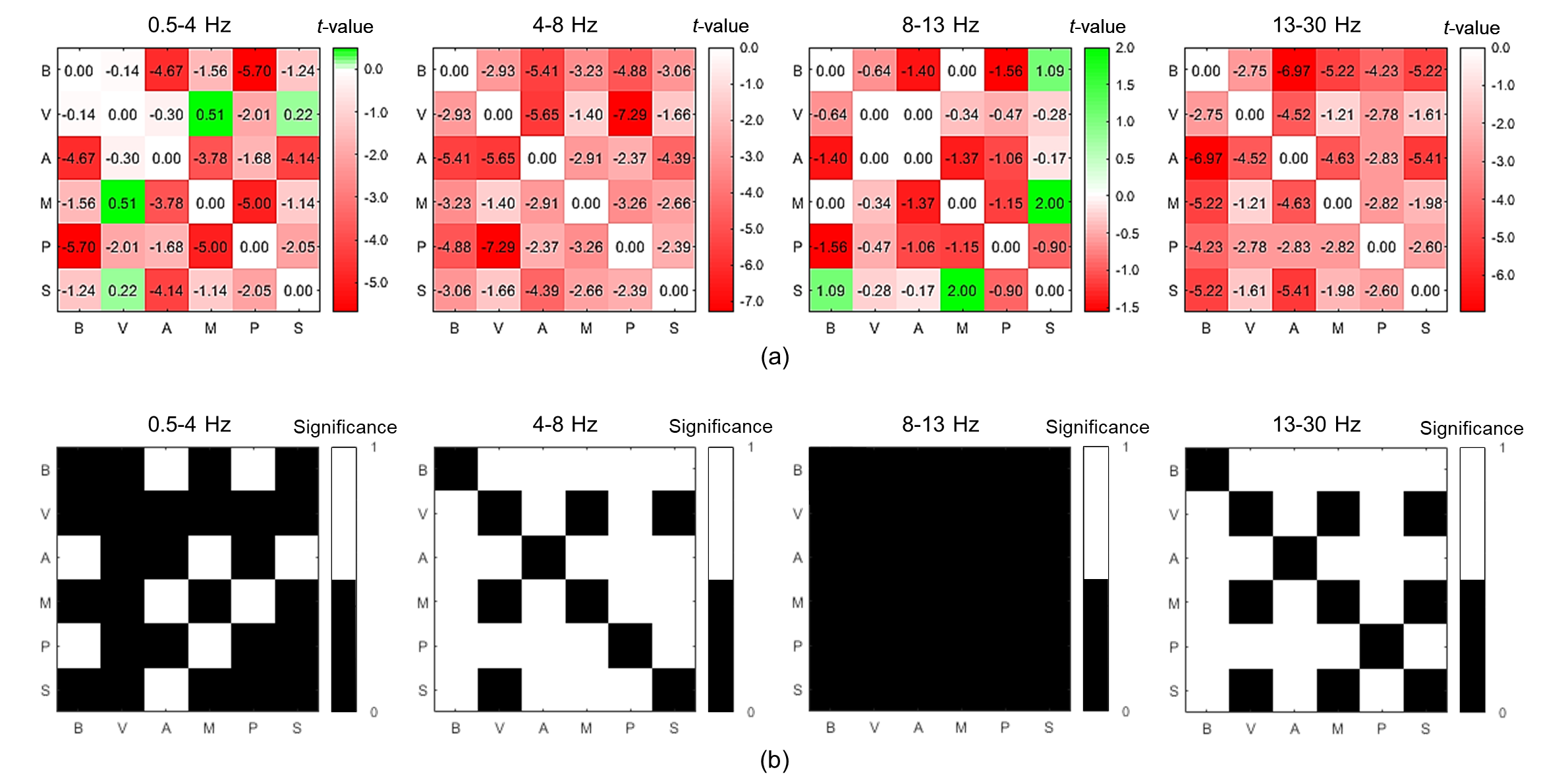}
\caption{Comparison between PLV of imagined speech and resting state. (a) The t-values of paired t-test  (B: Broca and Wernicke’s area; V: visual cortex; A: auditory cortex; M: motor cortex; P: prefrontal cortex; A: auditory cortex). (b) The corresponding significance of (a) from the paired t-test (white: p ${<}$ 0.05).}
\label{fig}
\end{figure*}

Fig. 1 displays the t-test results of comparing the grand-averaged inter-cortical PLV of imagined speech versus resting state. In the delta frequency region, the grand-averaged value of PLV between Broca and Wernicke’s areas and auditory cortex, Broca and Wernicke's area and prefrontal cortex, auditory cortex and motor cortex, auditory cortex and sensory cortex, and motor cortex and prefrontal cortex have displayed a significant decrease in PLV of imagined speech than the resting state. The inter-regional connectivity in the alpha band mostly exhibited a significant decrease among the cortical regions, except for the group between visual cortex to motor and visual cortex to sensory cortex. Similarly, the PLV of imagined speech significantly decreased between most regions, except for the visual cortex to motor cortex, visual cortex to sensory cortex, and motor cortex to sensory cortex. However, no significant decrease in the PLV of imagined speech was found between any cortical regions in the alpha frequency ranges. 

Similar aspects were shown in the case of visual imagery (Fig. 2). Delta connectivity displayed a significant decrease in every inter-cortical group except for Broca and Wernicke’s area to sensory cortex, visual cortex to motor cortex, visual cortex to sensory cortex; auditory cortex to motor cortex; auditory cortex to prefrontal cortex, auditory cortex to sensory cortex, prefrontal cortex to sensory cortex. In the theta band, a significant decrease in the PLV of visual imagery was found in Broca and Wernicke’s area to auditory, motor, prefrontal cortex; visual cortex to auditory and prefrontal cortex; motor cortex to prefrontal and sensory cortex; prefrontal cortex to sensory cortex. Alpha connectivity exhibited a significant decrease between Broca and Wernicke’s area to prefrontal cortex; visual cortex to auditory and prefrontal cortex; motor cortex to prefrontal cortex. Every inter-region connectivity of visual imagery in the beta band range displayed a significant decrease except for visual cortex to sensory cortex; motor cortex to prefrontal and sensory cortex; prefrontal cortex to sensory cortex.

\begin{figure*}[t!]
\includegraphics[width = \textwidth, height = 9cm]{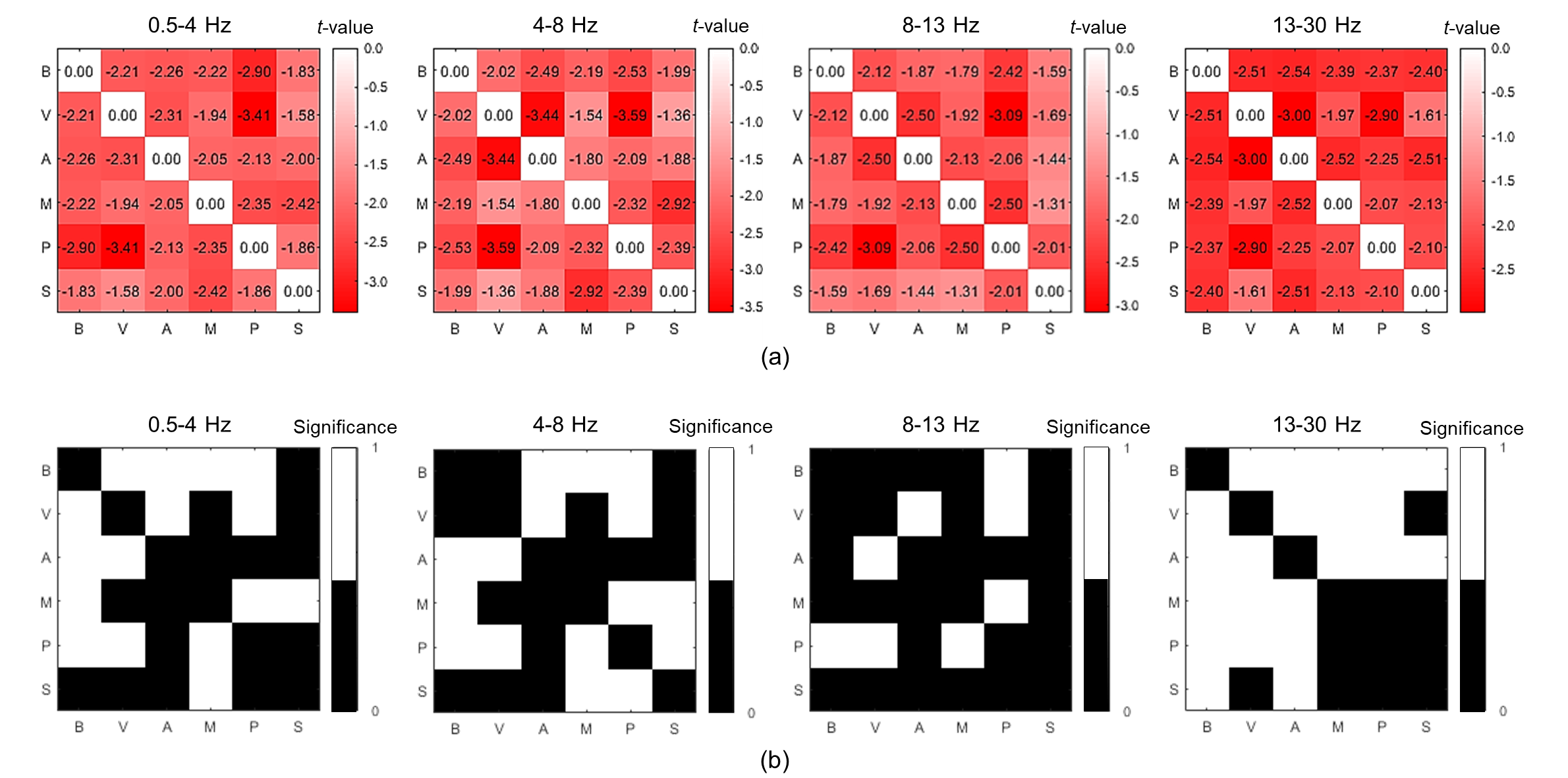}
\caption{Comparison between PLV of visual imagery and resting state. (a) The t-values of paired t-test  (B: Broca and Wernicke’s area; V: visual cortex; A: auditory cortex; M: motor cortex; P: prefrontal cortex; A: auditory cortex). (b) The corresponding significance of (a) from the paired t-test (white: p ${<}$ 0.05).}
\label{fig}
\end{figure*}

\section{Discussions}
\subsection{Decrease in PLV during Imagery}
During imagined speech and visual imagery, the PLV in the whole brain region significantly decreased, compared to the resting state. Similarly, a significant decrease in PLV while passing from the resting state to motor imagery (MI) has been reported in the previous works \cite{b22}. This is because the actual movement is inhibited during MI, resulting in the inhibited activation of the corresponding region \cite{b23}. The synchronization of the internal mental process is known to be opposed to the processing of external stimuli, leading to increased activation at rest and decreased activation during task performance \cite{b24}. Likewise, Nguyen \textit{et al}. \cite{b6} reported decreased brain activity during imagined speech. In line with these studies, a similar aspect may occur during imagined speech and visual imagery because they involve internal mental processes without other actions. This may be an important information in decoding the two paradigms, by focusing on the desynchronization during the active imagery.

\subsection{Functional Connectivity of Spectral Ranges}
The results of spectral analysis display a significant decrease in the frequency range mainly other than the alpha band, especially for the imagined speech. Since alpha power is involved with the relaxing state and is mainly found in the posterior regions, the active imagery may not have been reflected in the EEG of imagined speech. In the delta, theta, and beta band ranges, the PLV of both imagined speech and visual imagery has shown a significant decrease. The delta band is known to found during the continuous attention tasks and the beta band is known to be associated with active thinking, therefore, significance in the delta and beta band frequencies may be reasonable \cite{b25, n02}. Besides, since the subjects sat still without moving their body or speaking out, the inhibitory response may be reflected in the theta band  \cite{b25}. 

\subsection{Functional Connectivity of Cortical Regions}
The results on different cortical regions mainly exhibit a significant decrease in the imagined speech on Broca and Wernicke’s areas along with the auditory cortex. Since previous literature consistently described the left posterior superior temporal cortex (Wernicke’s area) involved in the language process, our results may be supporting the previous literature. Also, fronto-temporal coherence was reported as a significant measure of functional connectivity in speech processing  \cite{b12}. In this manner, decreased PLV during imagined speech in the temporal region may act as a relevant measure of functional connectivity. For visual imagery, decreased PLV was found in the auditory and the prefrontal cortex. The significance found in the prefrontal cortex may be supported with the top-down information flow in parieto-occipital cortices which is known as a distinct feature of visual imagery \cite{b14}. Also, a relevant distinction between visual imagery and visual perception is significantly displayed as the activation of the temporal regions \cite{b15}, therefore, our results may be supporting the previous literature. 

\subsection{Limitations and Future Works}
The correlation between resting state connectivity and the classification performance may have the potential as a performance predictor in imagined speech and visual imagery \cite{n06}. In fact, MI performance can be predicted using the connectivity in the resting state because the brain network before MI affects the brain signals during MI \cite{b20}. In this manner, further investigation of the correlation between brain connectivity and classification performance may be another crucial issue. Our results indicate that the functional connectivity of imagined speech or visual imagery decreases, compared to the resting state. It may imply that subjects whose brain connectivity has already been reduced in the resting state are easier to reduce their PLV sufficiently to achieve good performance for imagery \cite{b26}. However, we did not compare the resting state before the experiments; thus, further research is required.

\section{Conclusion}
Understanding the brain connectivity of the two emerging intuitive BCI paradigms may play a crucial role in decoding the two paradigms. Further investigation on the correlation between the brain connectivity and the decoding performance may highly contribute to practical BCI. Although we used the phase-locking value as a measure of functional connectivity, it holds limitations of being affected by other factors such as volume conductions in the brain. Therefore, investigation on the functional connectivity using other factors may support our findings. Also, further analysis on the effective connectivity of imagined speech and visual imagery may describe the directions of the connections, therefore, contribute in a thorough understanding of the brain dynamics that alter during the imagery of the two paradigms. Likewise, investigation of intuitive BCI paradigms along with the conventional paradigms \cite{n03, n04, n05} may ultimately improve the practicality of BCI works.




\vspace{12pt}

\end{document}